\documentclass[9pt,shortpaper,twoside,web]{ieeecolor}

\usepackage{generic}
\usepackage{cite}
\usepackage[T1]{fontenc}
\usepackage{float}
\usepackage{times}

\usepackage{soul}
\usepackage{url}
\usepackage[hidelinks]{hyperref}
\usepackage[utf8]{inputenc}
\usepackage[small]{caption}
\usepackage{graphicx}
\usepackage{amsmath}
\usepackage{amsfonts}
\usepackage{booktabs}
\usepackage{multirow}
\usepackage{subcaption}
\usepackage{tabularx}
\usepackage{xcolor}
\usepackage{wrapfig}
\usepackage{adjustbox}
\usepackage{textcomp}
\definecolor{ForestGreen}{RGB}{34,139,34}
\def\BibTeX{{\rm B\kern-.05em{\sc i\kern-.025em b}\kern-.08em
    Tkern-.1667em\lower.7ex\hbox{E}\kern-.125emX}}
\begin{document}
\title{FCSN: Global Context Aware Segmentation \\\ by Learning the Fourier Coefficients of Objects in Medical Images}
\author{Young Seok Jeon, Hongfei Yang, Mengling Feng
\thanks{Manuscript received xx xx, 2022; revised xx xx, xxxx , ... , and xx xx, xxxx; accepted xx xx, xxxx. Date of publication xx xx, xxxx; date of current version xx xx, xxxx.}
\thanks{This research is supported by the National Research Foundation Singapore under its AI Singapore Programme (Award Number: AISG-GC-2019-001 and AISG-GC-2019-002) and the NMRC Health Service Research Grant (MOH-000030-00).}
\thanks{YS. Jeon, H. Yang, M. Feng are with the Saw Swee Hock School of Public Health and Institute of Data Science, National University of Singapore, Singapore (e-mail: youngseokjeon74@gmail.com; hfyang@nus.edu.sg; ephfm@nus.edu.sg).}
\thanks{Equal contribution: Young Seok Jeon and Hongfei Yang.}
\thanks{Corresponding author: Mengling Feng (ephfm@nus.edu.sg).}
}

\maketitle

\begin{abstract}
The encoder-decoder model is a commonly used Deep Neural Network (DNN) model for medical image segmentation.
Conventional encoder-decoder models make pixel-wise predictions focusing heavily on local patterns around the pixel. This makes it challenging to give segmentation that preserves the object's shape and topology, which often requires an understanding of the global context of the object.
In this work, we propose a Fourier Coefficient Segmentation Network~(FCSN)---a novel DNN-based model that segments an object by learning the complex Fourier coefficients of the object's masks.
The Fourier coefficients are calculated by integrating over the whole contour. Therefore, for our model to make a precise estimation of the coefficients, the model is motivated to incorporate the global context of the object, leading to a more accurate segmentation of the object's shape.
This global context awareness also makes our model robust to unseen local perturbations during inference, such as additive noise or motion blur that are prevalent in medical images.
When FCSN is compared with other state-of-the-art models (UNet+, DeepLabV3+, UNETR) on 3 medical image segmentation tasks (ISIC\_2018, RIM\_CUP, RIM\_DISC),
FCSN attains significantly lower Hausdorff scores of 19.14 (6\%), 17.42 (6\%), and 9.16 (14\%) on the 3 tasks, respectively.
Moreover, FCSN is lightweight by discarding the decoder module, which incurs significant computational overhead.
FCSN only requires 22.2M parameters, 82M and 10M fewer parameters than UNETR and DeepLabV3+.
FCSN attains inference and training speeds of 1.6ms/img and 6.3ms/img, that is  8$\times$ and 3$\times$ faster than UNet and UNETR.
\end{abstract}


\section{Introduction}

Over recent years, we have witnessed increasing popularity in the applications of Deep Neural Network~(DNN) for various medical image segmentation tasks.
The encoder-decoder model~\cite{ronneberger2015u,chen2018encoder} is currently the most widely adopted DNN approach for the segmentation task.
Given enough training data, the encoder-decoder models can extract local patterns from an image that are associated with labels at each spatial coordinate.
However, due to its heavy reliance on local patterns, the model often fails to exploit the global contexts that potentially help to nullify nuisance local variations.

\begin{figure}[t!]
    \centering
    \includegraphics[width=\columnwidth]{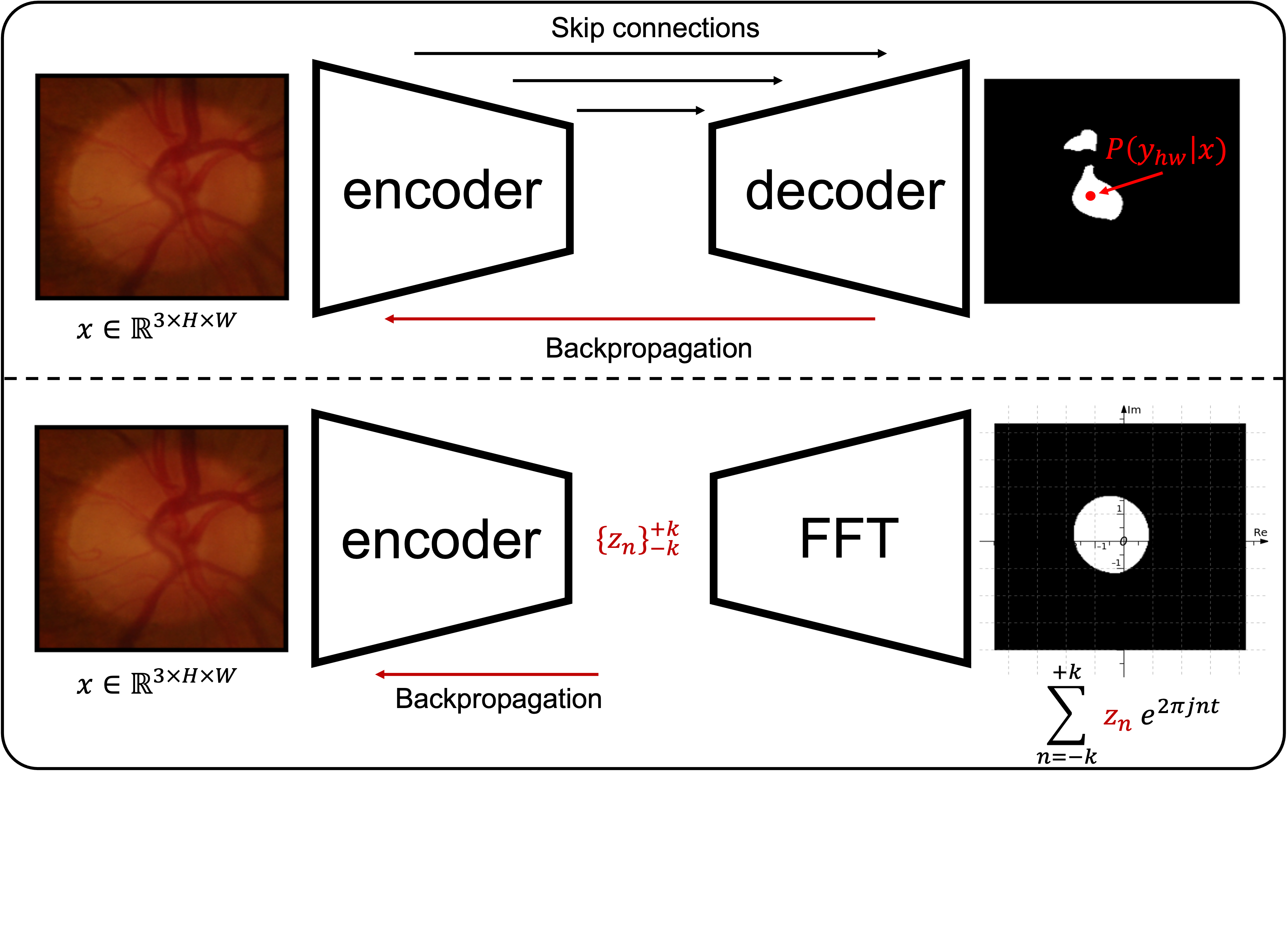}
    \vspace*{-12mm}
    \caption{Comparison of encoder-decoder model~(\textbf{upper}) and Fourier Coefficient Segmentation Network~(FCSN)~(\textbf{lower}). Unlike the encoder-decoder model, which makes a coordinate-wise prediction of an object, our FCSN predicts the complex Fourier coefficients of the object’s masks, which requires the learning of broader contextual information. Moreover, FCSN is more memory-efficient with the absence of a decoder.}
    \label{fig:overview}
\end{figure}

Specifically, in medical imaging tasks where the risk of misclassification is high, we need a model that is robust to many unpredictable local variations by incorporating the global contexts.
Taking the segmentation of optic cup in retinopathy as an example which is demonstrated in figure~\ref{fig:overview}, the following problems are difficult to address unless the model learns the global context: 
\begin{itemize}
    \item anatomically, the shape of an optic cup is always like a single filled oval, but current DNN often give segmentation with multiple components or with holes
    \item an optic disc has a smooth contour, but current DNN give contours with sharp corners or unnecessary zigzags
    \item retinopathy images from different sources are likely to suffer from different degradations, which cause generalization problems for current DNNs.
\end{itemize}

In this paper, we argue that these problems, which are either ignored or indirectly treated in the conventional encoder-decoder segmentation models, can be effectively addressed if we train the DNN to directly predict the shape, size and location of an object.

\subsection{Encoder-decoder Segmentation Model}

As shown in the first row of figure~\ref{fig:overview}, modern segmentation models typically adopt an encoder-decoder structure which models a conditional probability of predicting label $y_{hw}$ given an input $\mathbf{x}$ at each spatial coordinate $h, w$~($i.e.$ $\text{p}(y_{hw}|\mathbf{x})$).
The model is then optimized to maximize the likelihood of the spatially summed log probability ($i.e.$ $\text{argmax}_{\text{p}} \sum_{hw}  y_{hw}\log\text{p}(y_{hw}|\mathbf{x})$), assuming spatial independence across the coordinates.
Based on the structure of the model and the way in which the model is optimized, the existing encoder-decoder model will make a prediction mainly relying on local patterns and often does not utilize the global context of the image at all.
This absence of global context can cause inconsistency in segmentation performance, especially for the tasks that assume specific global priors.
Most of the existing works on global context learning aim to solve the problem by proposing a more flexible~(general) model structure that offers the model an opportunity of capturing global patterns~\cite{dosovitskiy2020image,liu2015parsenet,farabet2012learning}.
However, offering the opportunity does not necessarily mean that the model will explore the new aspect of learning.
There is a possibility that the model will still focus on finding local shortcut evidence and hence fails to focus on the global evidences~\cite{geirhos2020shortcut}.
Also, when the network is trained under a data constraint, higher flexibility could negatively impact the model performance.
In this regard, we argue that increasing the model flexibility alone is an unstable solution to the global context learning problem.

\subsection{Contribution}

We propose a novel segmentation model---Fourier Coefficient Segmentation Network (FCSN) that lifts segmentation to a shape prediction task, where the shape is represented as Fourier coefficients.
As shown in figure~\ref{fig:overview}, FCSN perceives the segmentation mask as a smooth function in a complex domain, which can be accurately approximated as complex Fourier coefficients.
We use Fourier Transform to extract the Complex Fourier coefficients of the contour of the mask.
Hence, FCSN learns the global shape of an object by predicting its Fourier coefficients, and during inference, a contour is retrieved with Inverse Fourier Transform.

To motivate how predicting Fourier Coefficients helps to learn global context, imagine we want to segment an ellipse-shaped object, which can be precisely described by three complex Fourier Coefficients $z_{-1},z_0, z_{1}$.
The $z_0$ describes the center of the ellipse, and $z_{-1}$ and $z_1$ determine lengths and orientations of the semi-major and semi-minor axes.
Thus, for a DNN to make a precise prediction of the three coefficients, the model must learn to perceive the whole ellipse as a single object.
This is in contrast to the traditional encoder-decoder model, where the model makes predictions only by looking at the local structure of the object.

Also, we propose to add a Fourier differentiable spatial to numerical transform (F-DSNT) module~\cite{nibali2018numerical} to improve the accuracy of Fourier coefficient prediction and also to reduce memory consumption.
One could view the coefficient prediction as a typical regression problem and introduce fully-connected (FC) layers on top of the spatially flattened feature.
However, FC layers have several drawbacks: 1) they are over-parameterized, affecting the generalizability, 2) it assumes a fixed input shape, and 3) the output range is not bounded.
Instead, DSNT drives the encoder module to produce heatmaps that represent the probability distributions of Fourier coefficients.
DSNT does not introduce any trainable parameter and works with any input shape.

We evaluate the performance of FCSN on three Medical image segmentation tasks, including skin lesion, optic disc, and optic cup segmentations. 
FCSN outperforms state-of-the-art segmentation models such as DeepLab-v3+ and U-Net+ when eveludated with Hausdoff Distance.
Furthermore, as our model can attend to global features, its performance does not degrade from local perturbations such as contrast change, additive noise, or motion blur.
Lastly, our model is lightweight, requiring less computational cost by discarding the decoder module that has been indispensable in the modern segmentation model and incurs a considerable memory overhead.

\begin{figure*}[hbt!]
    \centering
    \includegraphics[width=\textwidth, height=7.5cm]{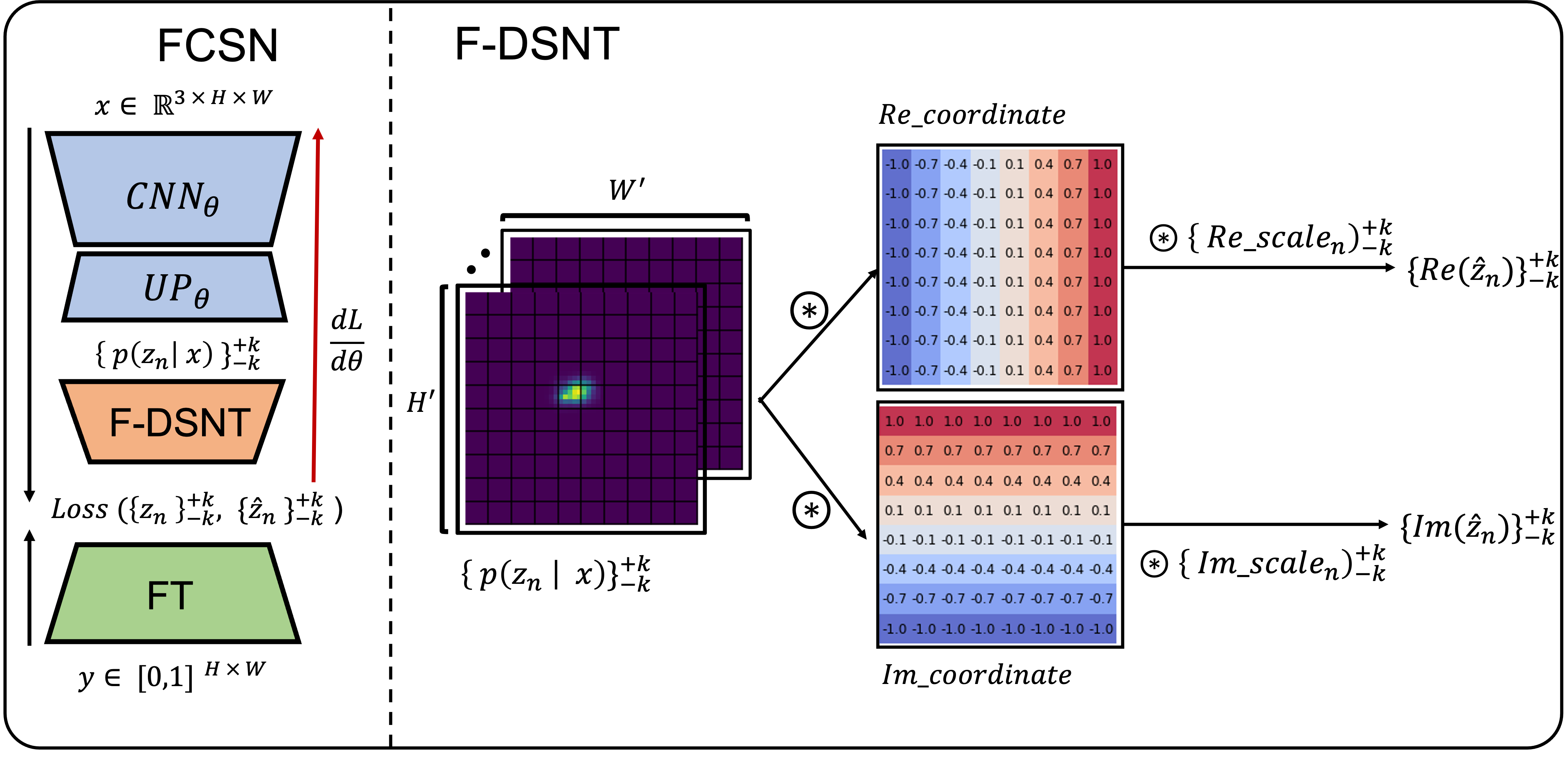}
    \caption{Overview of the FCSN architecture (\textbf{left}) and differentiable spatial to numerical transform (DSNT) (\textbf{right})}
    \label{fig:FCSN_overview}
\end{figure*}

\section{Related Work}

\subsection{Encoder-decoder Models}
FCN~\cite{long2015fully} and U-Net~\cite{ronneberger2015u} were the early few DNN models that proposed encoder-decoder structure for semantic segmentation.
However, the two approaches often produced noisy predictions that contained holes or non-smooth contours, implying that the models failed to understand the global context.
The issue had been addressed broadly in two ways while preserving the encoder-decoder structure: by 1) increasing the receptive field size and 2) introducing a regularizer that penalizes non-smooth prediction.

\subsubsection{Broader Receptive Field}
For a unit in the prediction of a network, the theoretical receptive field (TRF) of this unit refers to the region in the input image that contributes to the prediction of this unit. For convolution neural networks, the TRF is usually only a fraction of the input image, which depends on the architecture and filter sizes of the networks. To make more global aware predictions, the TRF must be large enough to cover the whole region that contains information related to the prediction.


In the literature several methods have been proposed to increase TRF. In \cite{liu2015parsenet} the authors proposed ParseNet which incorporated a global context feature that is generated using a global pooling operation in feature embedding.
In \cite{wang2020non,chen2021transunet} the authors proposed non-local U-Nets which included Transformer modules~\cite{vaswani2017attention} to extract long-range features.
In \cite{chen2018encoder} the authors proposed DeepLab with Atrous Convolution module that extracts features with varying receptive field sizes using dilated convolution.

As observed in \cite{luo2016understanding}, the effective receptive field (ERF) can be very different from theoretical receptive field. The ERF is defined as the collection of pixels inside TRF that have non-negligible impact on the prediction. It is found in \cite{luo2016understanding} that for neural networks prior training, the ERF is usually smaller than TRF, and a proper training is needed to enlarge ERF. Therefore, models with large TRF may not be capable of effectively understanding global context. In \cite{berman2018lovasz}, the authors proposed the Lov{\'a}sz metric, which is a convex function that approximates the Intersection over Union (IoU) metric. Since IoU is calculated over the whole image, the proposed metric can facilitate global learning.

\subsubsection{Regularizing Prediction}
Another approach to promote smooth segmentations is to adopt regularization on the models or the predicted masks. In~\cite{oktay2017anatomically} the authors proposed the ACNN-Seg for predicting high resolution segmentation masks from low resolution images. They introduced an extra autoencoder (AE) network to regulate segmentation outputs, such that the AE would produce similar features for both the predicted masks and the ground-truths.

%
More recently, the authors in~\cite{jia2021regularized,jia2020nonlocal} proposed to add spatial regularization to softmax activation functions in order to minimize total variation of predictions, such that the predicted masks are more robust to various local perturbations in the images.

\subsection{Segmentation via Shape}
For image segmentation most DNNs make per-pixel predictions for segmentation masks. One way to obtain more regularized prediction is to predict the shape of the segmentation mask, which effectively reduce the output dimensionality and complexity.

In \cite{liu2020abcnet,chen2021bezierseg} the authors proposed DNNs that learn parametrization of boundary curves via piecewise B\'ezier curves.
However, the B\'ezier parametrization does not necessarily converge to the true boundary curve.
In \cite{xie2020polarmask}, the authors proposed to predict polar coordinates of sampled points on boundary curves for instance segmentation.

There are not many DNN approaches that utilize Fourier transforms for segmentation.
In \cite{riaz2021fouriernet}, the authors used DNN to learn Fourier coefficients of sampled points on boundary curves for instance segmentation.
However, they regarded the $x$ and $y$ coordinates of boundary points as two sequences of real numbers and applied Fourier transforms independently.
In our approach, we regard the boundary curve as a sequence in the complex domain, and we apply Complex Fourier Transform only once to get Fourier coefficients.

\section{Proposed Method}

As shown in figure~\ref{fig:FCSN_overview}, our DNN model consists of four modules.
The first module $\text{CNN}_{\theta}$ is a feature extraction module that takes an image as its input. Any standard CNN backbone can be adopted.
The second module $\text{UP}_{\theta}$ generates heatmaps which represent the discrete probability distribution functions (PDF) of Fourier coefficients.
The third module $\text{F-DSNT}$ ``softly" picks up the most probable Fourier coefficient from each of the PDFs.
The last module $\text{FT}$ recovers segmentation masks from the predicted Fourier coefficients. To understand our approach, we explain how we convert masks to Fourier coefficients first.

\subsubsection{\textbf{FT} : Segmentation Masks to Fourier Coefficients}

Let $Y$ be a binary segmentation mask.
We regard $Y$ as a function on the complex domain $D=\{x+jy:-1\leq x,y\leq 1\}$, where $Y(x+jy)=1$ for foreground and  $Y(x+jy)=0$ for background.
Let $\alpha:[0,1]\to \mathbb C$ be a parametrization of the boundary curve of foreground. We assume $\alpha$ is a complex valued smooth curve with $\alpha(0)=\alpha(1)$. Given the boundary curve $\alpha$, the region enclosed by $\alpha$ is the segmentation region.

The Fourier coefficients $\{z_n\ \in \mathbb{C}\}$ of the boundary curve $\alpha(t)$ is defined by

\begin{equation}
\label{eq:Fourier_eqn}
    z_n = \int_0^{1} \alpha(t)e^{-2\pi jnt}\,\mathrm{d}t
\end{equation}
for $n= \dots,-1,0,1,\dots$, where $j$ is the imaginary unit.
The original boundary curve $\alpha$ can be fully recovered from the Fourier coefficients $\{z_n\}$ by taking the Inverse Fourier transform defined by

\begin{equation}
    \alpha(t) = \sum_{n=-\infty}^{\infty} z_n e^{2\pi jnt}.
\end{equation}
Therefore, instead of making a direct prediction of the segmentation mask $Y$, it is possible to predict the Fourier coefficients $\{z_n\}$ and recover the mask $Y$ with Inverse Fourier transform.

\begin{figure}[t]
    \centering
    \includegraphics[width=5 cm]{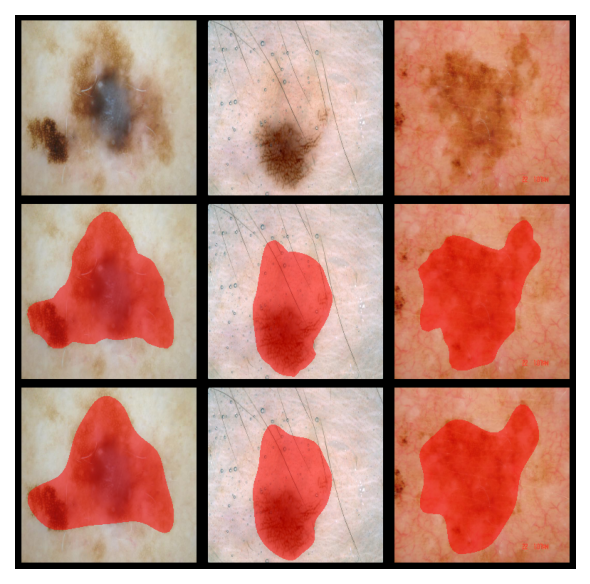}
    \caption{Original mask (second row) and masks generated from boundary curves with $21$ Fourier coefficients (third row).}
    \label{fig:Fourier_low}
\end{figure}

Predicting Fourier coefficients forces the training of DNN to utilize global context better.
As suggested by equation \eqref{eq:Fourier_eqn}, the Fourier coefficients, which we predict, are obtained by integrating global information on the boundary curve.
This forces DNN models to learn the global context of an image better, facilitating to make more spatially consistent segmentation.

It is usually sufficient to only learn to predict the lower Fourier coefficients which encodes the location and the general shape of the boundary curve $\alpha$. This is because the coefficients $\{z_n\}$ are concentrated on small absolute values of $n$ when $\alpha$ is smooth: In fact, if $\alpha$ is $k$-times continuously differentiable, then $z_n$ converges to $0$ faster than $1/|n|^k$ for large $n$.
Discarding higher Fourier coefficients can be regarded as a regularization that smooths ground-truth boundary curves.
Figure~\ref{fig:Fourier_low} shows segmentation masks obtained by only taking $z_n$ for $-10\leq n\leq 10$.

\subsubsection{$\mathbf{UP_{\theta}}$ : Probability Distribution of Coefficients}
Given a feature extracted from a raw input using a CNN module, $\text{UP}_{\theta}$ generates heatmaps which represent the discrete PDFs of possible Fourier coefficients. ($i.e.$ $\{ \text{p}(z_n | \mathbf{x}) \}_{-k}^{+k} = \text{UP}_{\theta} \circ \text{CNN}_{\theta} $).
$\text{UP}_{\theta}$ module consists of a 2D transposed convolution layer with $2*k + 1$ kernels, followed by a softmax activation across spatial axes.
2D transposed convolution layer projects input feature to a higher spatial resolution; thus, the generated heatmaps are more granular.
We apply softmax to normalize the heatmaps such that it is non-negative and sum to one.

\subsubsection{$\mathbf{F-DSNT}$ : Selecting the Most Probable Coefficients}

Finding the most probable coefficient from each discrete PDF ($i.e.$ $\hat{z}_n = \text{argmax} \;  \text{p}(z_n | \mathbf{x})$) is not differentiable.
To make it differentiable, we adopt DSNT~\cite{nibali2018numerical} which can be viewed as a soft-argmax operation.
This is done by calculating the expectations of the PDFs.
As shown in figure~\ref{fig:FCSN_overview}, the expectations are calculated by performing a weighted sum of discrete PDF with real and imaginary coordinate values.
%



For the original implementation of DSNT in \cite{nibali2018numerical}, the PDFs are assumed to have spatial range $[-1,1]\times [-1,1]$. 
In our model, we multiply the output of our DSNT module with scaling constants estimated by checking the range of each Fourier coefficient from the training dataset.
This is equivalent to increasing the resolution of PDFs for higher Fourier coefficients which are usually close to zero.

\subsubsection{Loss Function}

Our loss function is a combination of weighted $L_1$ and $L_2$ losses plus the Jensen-Shannon (JS) divergence regularization. 
Given a batch of $M$ input images $\{\mathbf{x}^{(m)}\}$, our predicted coefficients $\{\hat z_n^{(m)}:-k \leq n \leq k\}$, and the ground truth Fourier coefficients
$\{z_n^{(m)}:-k \leq n \leq k\}$,  the loss function is 

\begin{equation}
\begin{aligned}
 \text{Loss}(z_n, \hat z_n) = \frac{1}{M}\sum_{m,n} \Big\{ & w_n \left( |\hat z_n^{(m)}- z_n^{(m)}| + |\hat z_n^{(m)}- z_n^{(m)}|_2^2 \right) \\
 & + \text{JS}(p(\hat z_n^{(m)}|\mathbf{x}^{(m)})||\mathcal{N}(\hat z_n^{(m)},\sigma I_2)) \Big\},
\end{aligned}
\end{equation}
where $p(\hat z|\mathbf{x})$ is the PDF generated by our $\text{UP}_{\theta}$ module.
The $w_n$'s are weight constants that we introduce to promote the learning of higher Fourier coefficients which are much smaller than lower coefficients, defined as 
\[
w_n = \min \left\{1+\frac{1}{\max_i | z_n^{(i)}|+\varepsilon},10\right\}.
\]
The $\text{JS}(p(\hat z|\mathbf{x})||\mathcal{N}(\hat z,\sigma I_2))$ is the JS divergence between the PDF $p(\hat z|\mathbf{x})$ and the bivariate normal PDF $\mathcal{N}(\hat z,\sigma I_2))$ with the same mean. The covariance $\sigma$ of the bi-normal PDF is a hyper parameter. The JS regularzation is minimized when the heatmap matches with the Gaussian distribution, thus making sure our heatmaps of Fourier coefficients are unimodal and concentrate nicely around true locations of the Fourier coefficients.

\section{Experiments}

\subsection{Evaluation Metrics}

\begin{figure}[h]
    \centering
    \begin{subfigure}[t]{0.1\textwidth}
        \centering
        \includegraphics[height=0.65in]{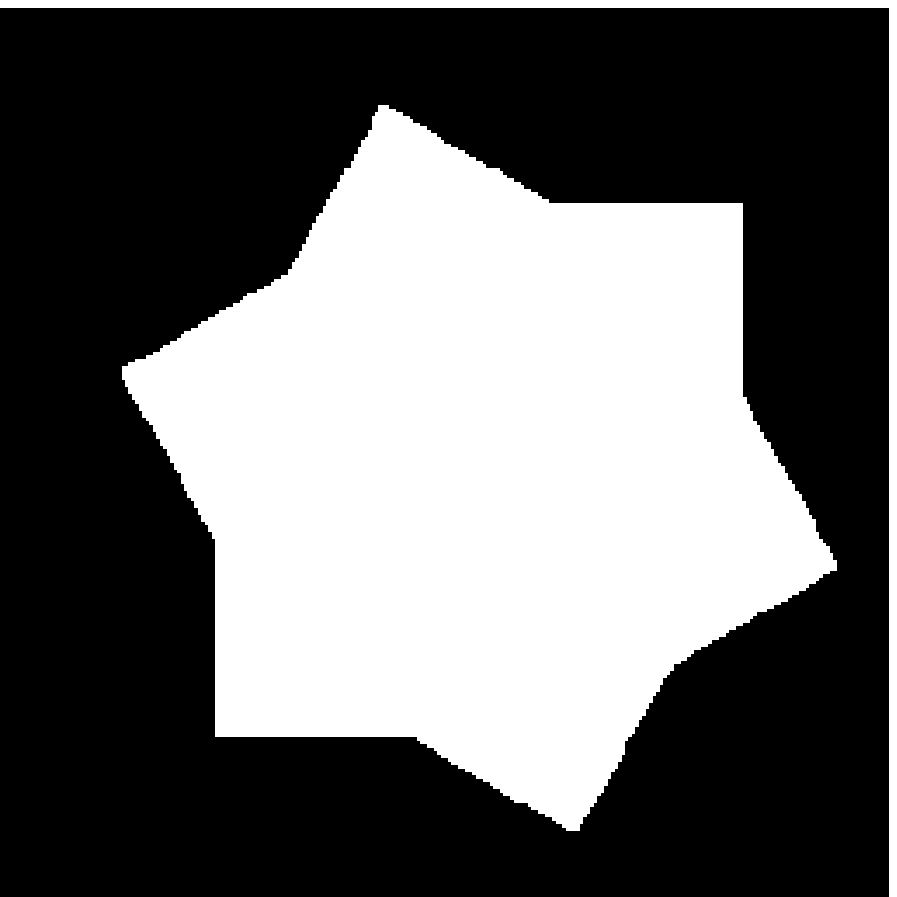}
        \caption{}
    \end{subfigure}%
    ~ 
    \begin{subfigure}[t]{0.1\textwidth}
        \centering
        \includegraphics[height=0.65in]{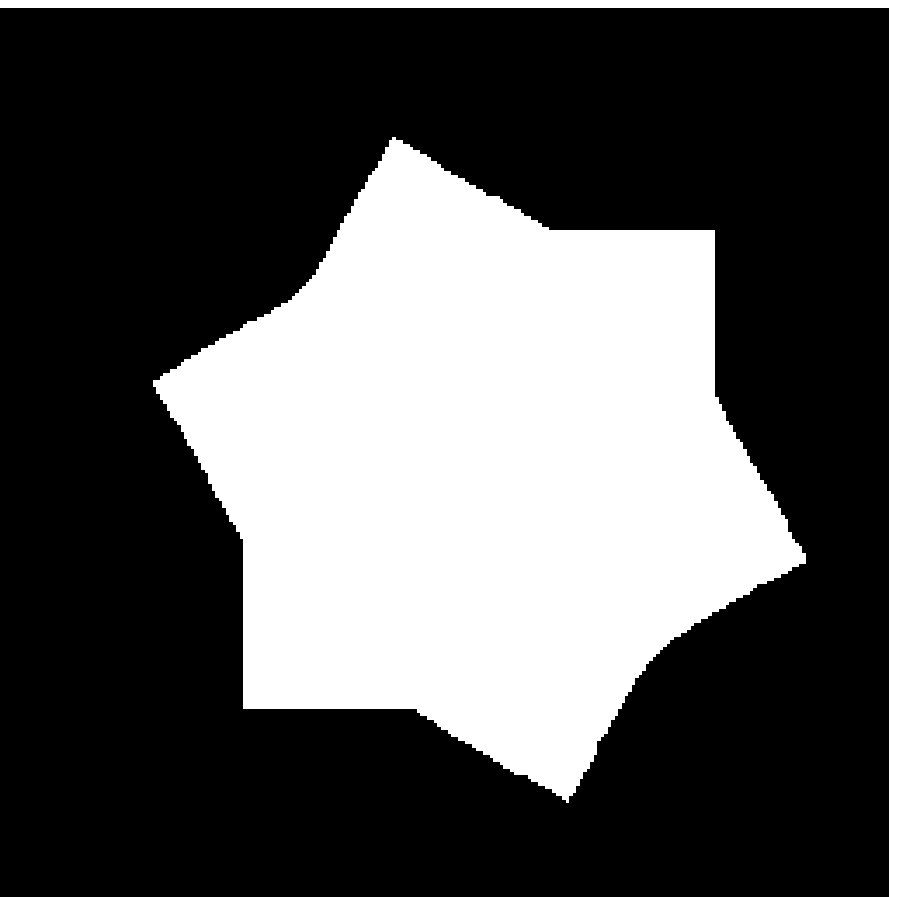}
        \caption{}
    \end{subfigure}
    ~ 
    \begin{subfigure}[t]{0.1\textwidth}
        \centering
        \includegraphics[height=0.65in]{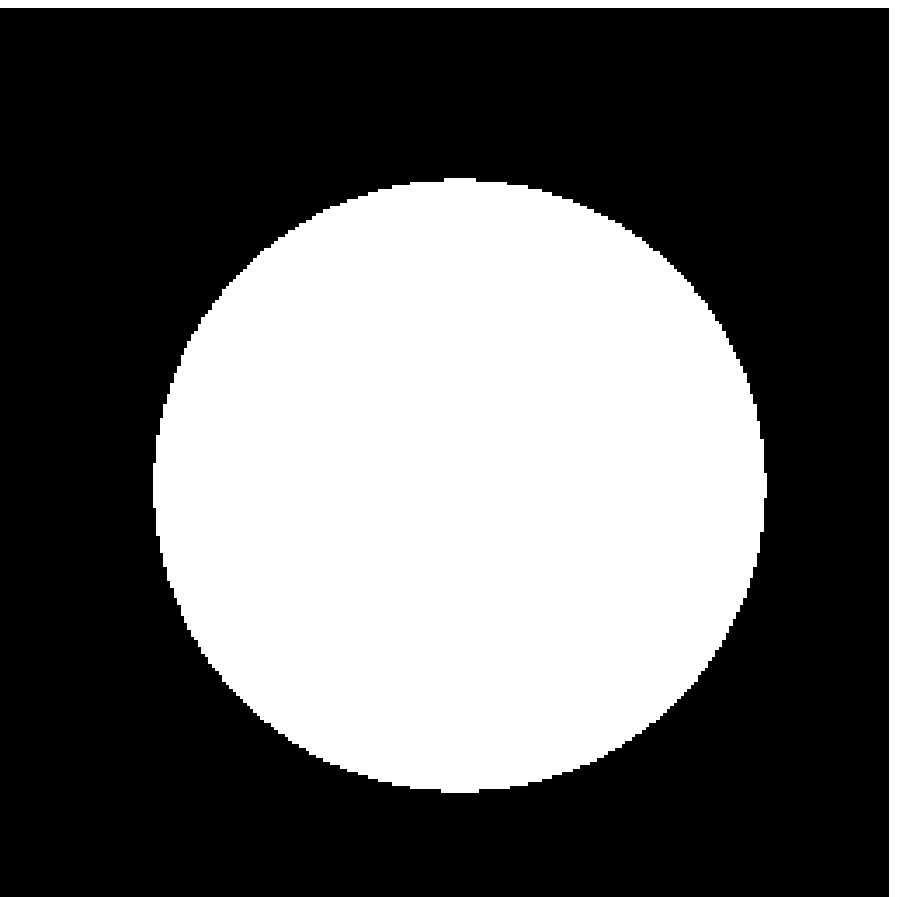}
        \caption{}
    \end{subfigure}
    ~ 
    \begin{subfigure}[t]{0.1\textwidth}
        \centering
        \includegraphics[height=0.65in]{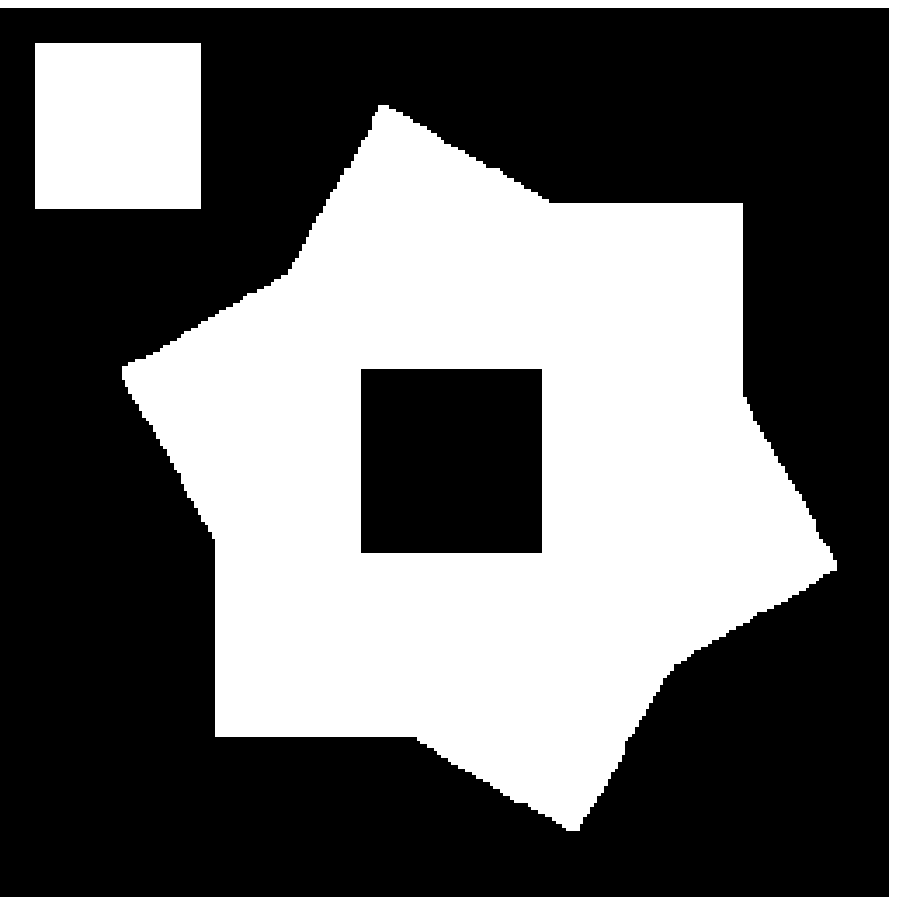}
        \caption{}
    \end{subfigure}
    \caption{(a) Ground truth, (b)--(d) three predictions with the same dice value $0.9$ but Hausdorff distances (smaller is better) $11.3$, $26.9$ and $93.3$ respectively. Note that the star shape in (b) is smaller than that in (a).}
    \label{fig:metrics_comparison}
\end{figure}

\begin{table*}[t]
    \caption{Dice \& Hausdorff comparison between FCSN and baseline encoder-decoder models on 3 tasks. The standard deviation (std) is computed from 5-fold results. The best result is in bold.}
    \centering
    \begin{adjustbox}{width=1\textwidth}
    \small
    \begin{tabular}{l cc cc cc c r}
        \toprule
        {} & \multicolumn{2}{c}{\bfseries ISIC\_2018} & \multicolumn{2}{c}{\bfseries RIM\_CUP} & \multicolumn{2}{c}{\bfseries RIM\_DISC} & {} & {} \\
        \cmidrule(l){2-3} \cmidrule(l){4-5} \cmidrule(l){6-7} 
        
        \bfseries Models & \bfseries Haus$\pm \text{std}$ & \bfseries Dice$\pm \text{std}$ & \bfseries Haus & \bfseries Dice & \bfseries Haus & \bfseries Dice & \bfseries \# Parameter (M) & \bfseries \# Flops (G) \\
        \midrule
        UNet         & 25.00 $\pm$ 1.00 & 0.89 $\pm$ 0.01 & 22.35 & \textbf{0.78} & 10.78 & 0.96 & 31.39 & 55.84 \\
        UNet+        & 24.06 $\pm$ 0.80 & 0.89 $\pm$ 0.01 & 22.25 & 0.77 & 11.79 & 0.96 & 36.63 & 138.16 \\
        DeepLabV3 (ResNet50)  & 20.79 $\pm$ 1.09 & $\mathbf{0.90 \pm 0.01}$ &  21.69 & 0.77 & 10.92 & 0.96 & 39.63 & 40.99 \\
        DeepLabV3+ (ResNet50)  & 20.80 $\pm$ 0.76 & $\mathbf{0.90 \pm 0.01}$ & 22.25 & 0.77 & 11.27 & 0.96 & 39.76 & 43.31 \\
        \midrule
        DeepLabV3+ (ResNet50 + Lov{\'a}sz) & 20.44 $\pm$ 1.34 & $\mathbf{0.90 \pm 0.01}$ & 20.60 & \textbf{0.78} & 10.63 & 0.96 & 39.76 & 43.31 \\        
        UNETR (VIT-B-16)        & 20.50 $\pm$ 0.76 & $\mathbf{0.90 \pm 0.01}$ & 18.98 & \textbf{0.78} & 10.95 & 0.96 & 105.32 & 32.14 \\
        \midrule
        FCSN (ResNet50)   & 20.21 $\pm$ 0.88 & 0.88 $\pm$ 0.01 & 18.15 & 0.77 & 9.85 & 0.96 & 29.71 & $\mathbf{23.54}$ \\
        FCSN (DResNet26)  & 20.14 $\pm$ 1.00 & 0.88 $\pm$ 0.01 & 18.07 & 0.77 & 9.59 & 0.96 & $\mathbf{22.16}$ & 83.01 \\
        FCSN (DResNet50)  & $\mathbf{19.14 \pm 0.86}$ & 0.88 $\pm$ 0.01 & $\mathbf{17.42}$ & $\mathbf{0.78}$ & $\mathbf{9.16}$ & 0.96 & 29.71  & 98.11 \\
        \bottomrule
    \end{tabular}
    \end{adjustbox}

    \label{table:accruacy}
\end{table*}

Let $Y$ be a segmentation mask, and let $\hat Y$ be a mask predicted by a DNN model. To measure model performance, we use both the dice metric and the Hausdorff distance defined by
\begin{equation}
    \label{eq:hausdorff_distance}
    \mathrm{H(Y,\hat Y)} = \max \left \{ \sup_{Y(y)=1}d(y,\hat Y), \sup_{\hat{Y}(y)=1}d(y, Y)\right\},
\end{equation}
where $d(y,Y)$ is the Euclidean distance from the point $y$ to the target in $Y$, and $d(y,\hat Y)$ is defined similarly. The smaller the Hausdorff distance is, the better the approximation of $\hat Y$ is to $Y$, and $\mathrm{H(Y,\hat Y)}=0$ means $Y$ and $\hat Y$ coincides completely.

The dice metric is widely used in evaluating segmentation models. However, the dice metric is not sensitive to changes of shape and topology of the masks. This is demonstrated in figure~\ref{fig:metrics_comparison}, where (a) is the ground truth, and (b)--(d) are three predictions with the same dice value $0.9$. However, it is clear that Figure~\ref{fig:metrics_comparison}(b) gives the best segmentation, while the shape of the segmentation in (c) is wrong, and the topology of the segmentation in (d) is wrong. On the other hand, the Hausdorff distance is more sensitive to changes in shape and topology, and it can successfully pick up the best segmentation.

\subsection{Datasets}
We use two publicly available medical image datasets: 1) ISIC-2018~\cite{codella2019skin} and 2)  RIM-ONE-DL~\cite{RIMONEDLImageAnalStereol2346}.

\subsubsection{ISIC-2018}

ISIC-2018 dataset contains 2,594 and 100 dermoscopic images with ground truth segmentation for training and validation, respectively.
The test dataset is not publicly available.
Hence, following conventions of other papers using ISIC, we report the final evaluation results using 5-fold cross-validation on the training dataset.

\subsubsection{RIM-ONE-DL}
    
RIM-ONE-DL dataset contsists of 313 and 172 retinographies from normal and glaucoma patients respectively.
All images include a manual segmentation of disc and cup that have been assessed by experts.
The dataset contains 341 and 149 training and testing samples respectively.
As suggested by the dataset provider, we perform a simple train-test split evaluation.

\subsection{Implementation Details}

During training and inference, images are resized to have a size $256\times 256$. 
For data augmentations, we used ColorJitter, random crop, and random flip for the RIM dataset, and we replaced random crop by resizing and random crop for the ISIC dataset.
For all our training, we trained for 500 epochs with a batch size of $8$, and we used the Adam optimizer~\cite{kingma2014adam} with a learning rate of $3e^{-4}$ without weight decay.
To generate Fourier Coefficients, we sampled $71$ points on boundary curves and used FFT to get the Fourier Coefficients, where the model only learns $21$ lower coefficients ($i.e.$  $\{z_n\}_{-10}^{+10}$). All our codes will be made publicly available later upon acceptance.

\begin{figure*}[t]
    \centering
    \includegraphics[width=15cm]{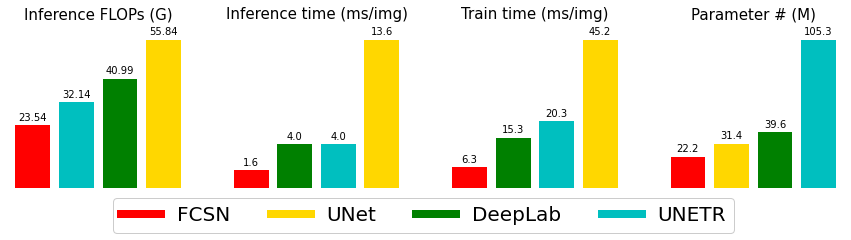}
    \caption{Computational efficiency of FCNS and baseline models in 4 different aspects:
Floating point operations per second (FLOPs), inference \& train speed (ms/img), and model size (M).}
    \label{fig:comlex}
\end{figure*}

\subsection{Results}

\subsubsection{Precise Shape Prediction}

We compare the performance of FCSN with different backbone settings against state-of-the-art segmentation models, including UNet \cite{ronneberger2015u}, UNet+ \cite{zhou1807nested} and DeepLab-v3+~\cite{chen2018encoder} with/without the lov{\'a}sz-softmax loss \cite{berman2018lovasz}, and UNETR~\cite{hatamizadeh2022unetr} (UNet with Transformer).
We perform experiments on 3 segmentation tasks: ISIC skin lesion, RIM\_CUP, and RIM\_DISC.
The model performance is accessed with Hausdorff and Dice metrics.

As shown in Table~\ref{table:accruacy}, for all instances, FCSN achieves a significantly lower Hausdorff score while maintaining a competitive Dice score, supporting that the shape of generated mask closely matches with ground truth.
Also, we observe greater performance gain from RIM\_CUP and RIM\_DISC tasks that have much smoother contours than ISIC.
We note that the performance of FCSN improves when we use DResNet~\cite{yu2017dilated} backbone that produces higher resolution output.
Also, using a deeper backbone further improves the performance.

\begin{figure}[h]
    \centering
    \includegraphics[width=\columnwidth]{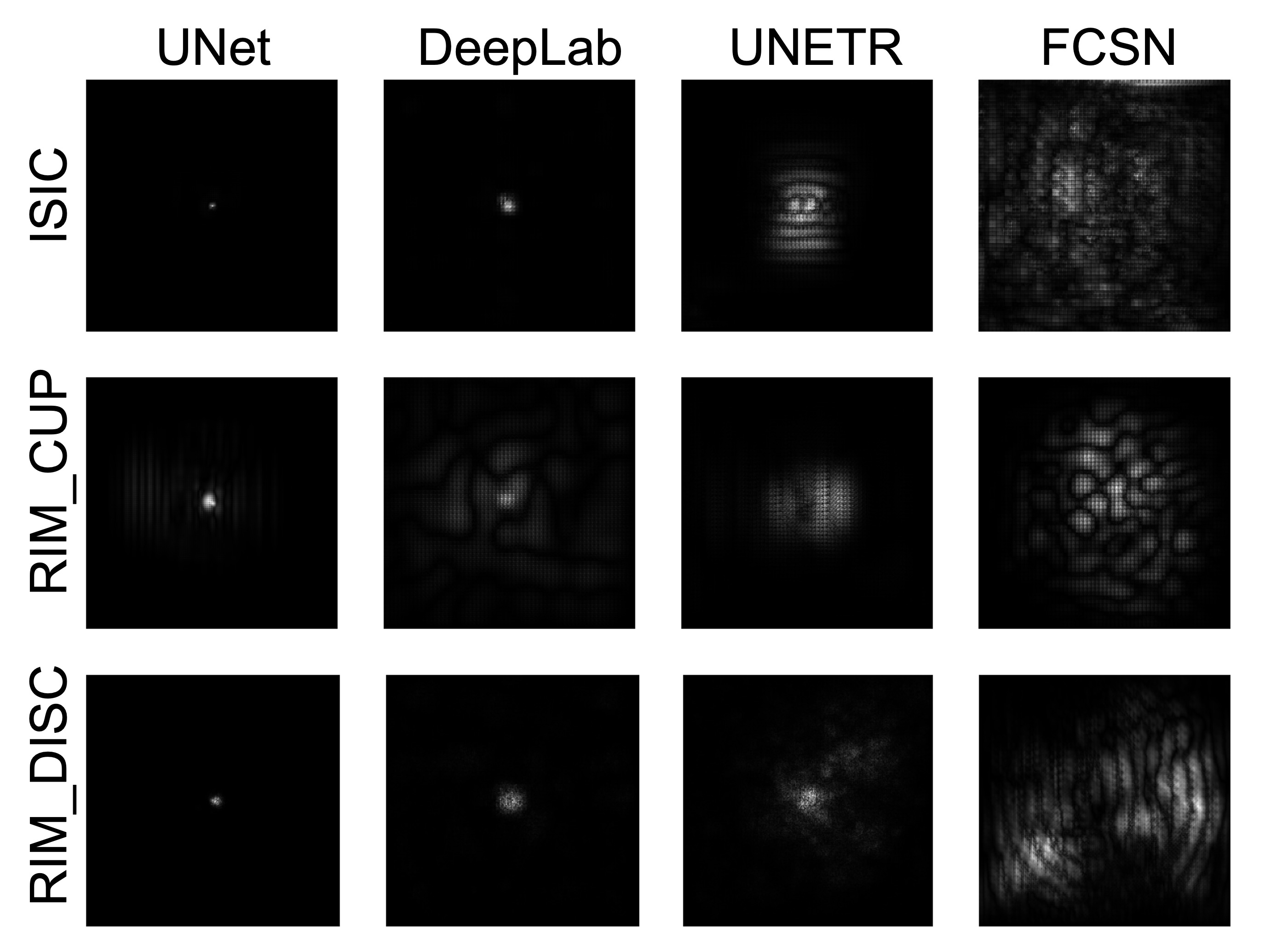}
    \caption{Comparison of Effective Receptive Field~(ERF) }
    \label{fig:erf}
\end{figure}

\subsubsection{Robustness to Perturbations}

\begin{figure*}[t]
\centering
\includegraphics[width=15cm]{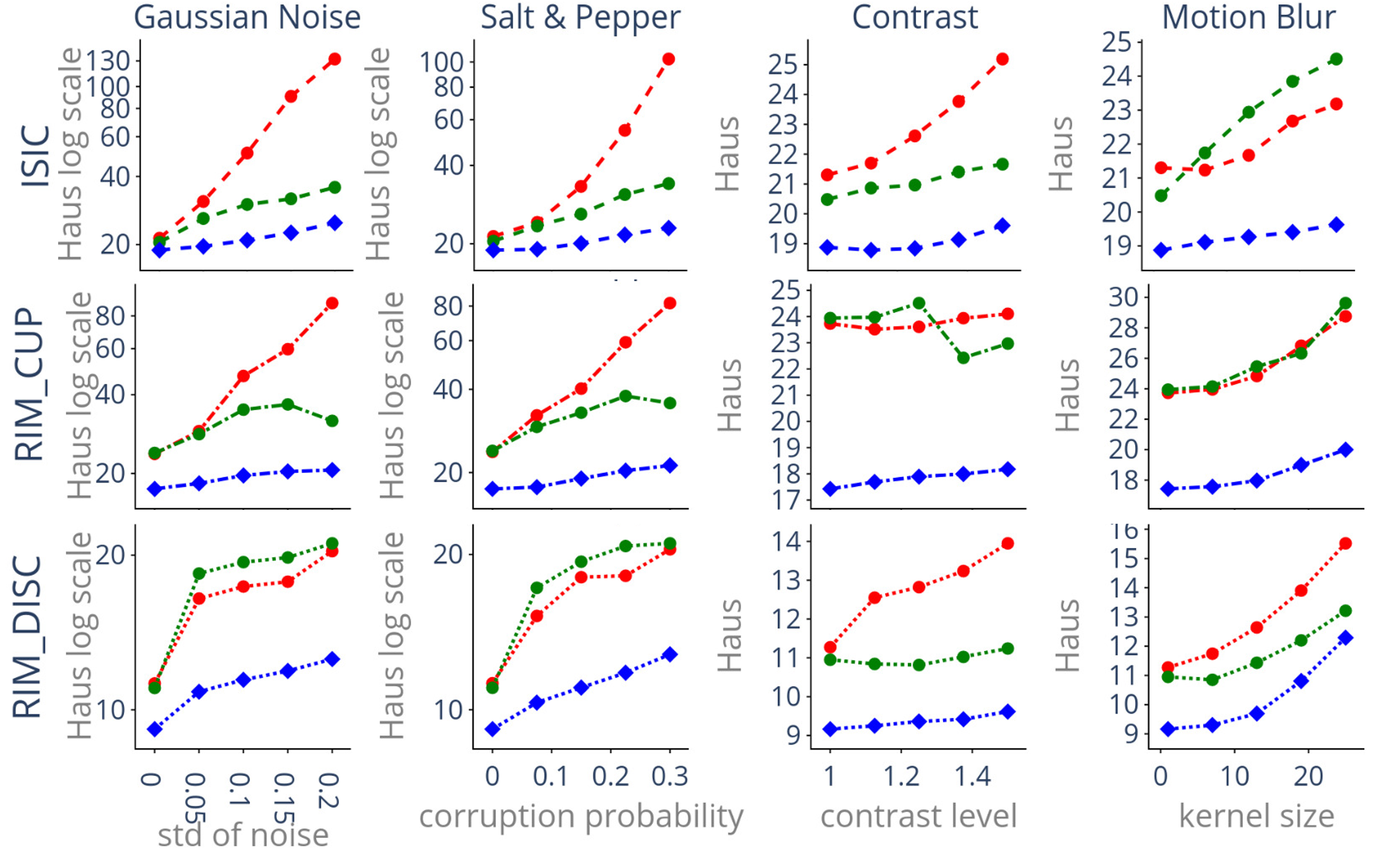}
\caption{Hausdorff distance (smaller the better) of inferences with perturbations. {\color{red} Red: DeepLab-v3+ (Lov{\'a}sz loss)}, {\color{ForestGreen} Green: UNETR}, {\color{blue} Blue: FCSN}. FCSN is more robust to perturbations, especially for heavy noises.}
    \label{fig:perturbation_test}
\end{figure*}

We test robustness of models to four types of perturbations at inference: Gaussian noise, Salt \& Pepper noise, contrast changes, and motion blur.
We chose Gaussian and Salt \& Pepper noises because they are the most common additive and impulsive noises respectively.
Contrast change and motion blur are typical degradation in medical images.
The results are summarized in figure~\ref{fig:perturbation_test}, where level of perturbation increases along the x-axis. Comparing with the DeepLab-v3+ (with Lov{\'a}sz loss) and the UNETR models, our method is more robust, especially for the two  noises, where our method can give almost consistent predictions regardless of noise level; on the other hand, the predictions of the DeepLab-v3+ model deteriorate heavily as noise level increases. Metrics of results of the UNETR model are either similar to that of DeepLab-v3+, or lie between the DeepLab-v3+ and our method.

\begin{figure*}[t]
    \centering
    \includegraphics[width=17cm]{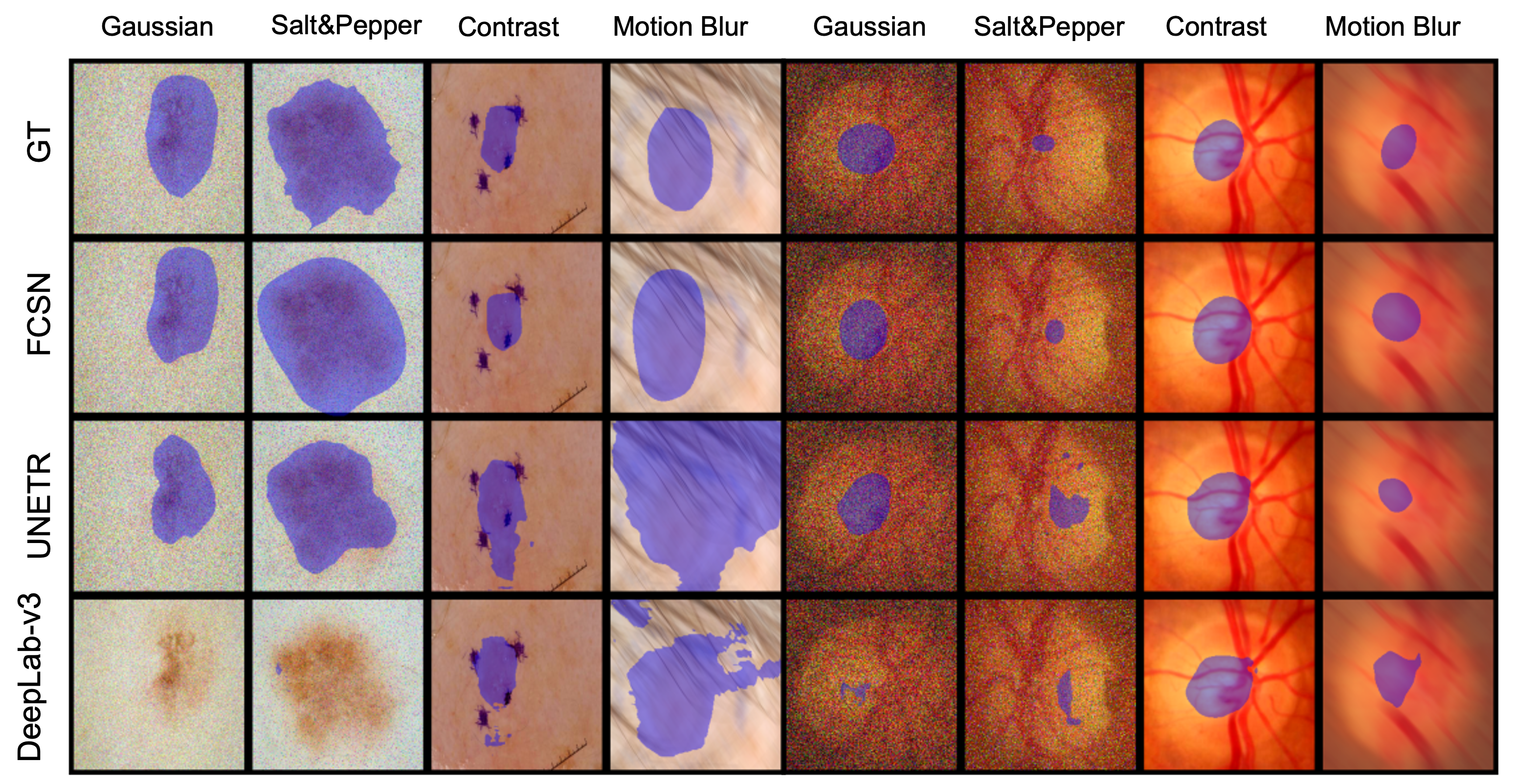}
    \caption{Visual comparison of predicted masks with perturbations.}
    \label{fig:noises}
\end{figure*}

Figure~\ref{fig:noises} shows examples of segmentation results for images with perturbations. For images with noise or contrast change, the DeepLab-v3+ method omitted large portions of target areas, and the UNETR failed to correctly segment the RIM cup with Salt \& Pepper noise, while our method consistently give reasonable segmentation for all cases. For the image with motion blur, the DeepLab-v3+ and UNETR methods wrongly included large portion of background area. All the predictions of the DeepLab-v3+ have either wrong shape or wrong topology. On the other hand, our method gives satisfactory segmentation results.

\subsubsection{Global Context Awareness}
Here, we empirically prove that the two major strengths of FCSN, precise shape prediction and robustness to perturbations, indeed arise from the model's global context awareness.
We propose to use the Effective Receptive Field (ERF), initially proposed by Luo et al.~\cite{luo2016understanding}, as the method to measure the global context awareness of models.
ERF measures how much each input pixel contributes to the model prediction.
Mathematically, this is done by computing the partial derivative of an arbitrary output unit $y_i$ with respect to input tensor $\mathbf{x}$ $\text{i.e} \; \partial y_i / \partial \mathbf{x}$, measuring how much $y_i$ changes as $\mathbf{x}$ changes by a small amount. ERF is therefore a natural measure of the importance of $\mathbf{x}$ with respect to $y_i$. 

Figure~\ref{fig:erf} shows the comparison of ERF for various models.
We observe that FCSN visually attains significantly bigger ERF size compared to baseline models across all tasks, strongly supporting our global context awareness argument.

\subsubsection{Computational Efficiency}

We compare the computational efficiency of FCSN against baseline segmentation models.
Specifically, we measure models' floating-point operations per seconds (FLOPs), inference time (ms/img), training time (ms/img), and parameter number (M).
We compare FCSN with ResNet50 backbone against UNet, DeeLab with ResNet 50 backbone, and UNETR with VIT-B-16 backbone. 
During the measure of FLOPs, inference \& training time, we set the input size to 256 $\times$ 256.
The results in Figure~\ref{fig:comlex} shows the computational efficiency of FCSN in all of the 4 aspects.
Comparing with the least performing model for each of the aspect,
FCSN requires 58\% less FLOPs, 8$\times$ faster training and inference speed, and 5$\times$ less parameter number. Note that the computation overheads from Fourier transform and inverse Fourier transform are small, which are equivalent to two 1d convolution layers with kernel size of 21 and input size 21. Empirically, these two transforms only take 0.05ms/img.

Our model has high computational efficiency because our model does not contain a conventional decoder. For most segmentation models employing neural network approach, they contain decoders which have several layers of 2d convolution and up-sampling operations. This will introduce a large amount of model parameters and heavy computations. On the other hand, our model only contains the encoder, and the prediction of Fourier coefficients is based on the F-DSNT layer, which incur little computation and does not contain learnable parameters.

\subsection{Ablation}
\subsubsection{Impact of DSNT}
For comparison, we remove $\mathbf{UP_{\theta}}$ and $\mathbf{DSNT}$ parts of our model and connect the feature maps from our backbone to FC layers to get Fourier coefficients.
Experiment results in table~\ref{table:ablation_DSNT} show that for the Dice metric, the DSNT approach consistently gives better results, while for the Hausdorff metric, the DSNT approach gives better results in most of the cases.


\begin{table}[t]
    \centering
    \resizebox{\columnwidth}{!}{
    \begin{tabular}{lllcccc}
        \toprule
        \multirow{2}{*}{} & \multirow{2}{*}{}  & \multirow{2}{*}{} & \multicolumn{4}{c}{\bfseries Epoch Number} \\
        \cmidrule(l){4-7} 
        {\bfseries Tasks} & {\bfseries Metric} & {\bfseries Heads} & 100 & 200 & 300 & 400 \\
        \midrule
        \multirow{4}{*}[-0.1em]{ISIC} & \multirow{2}{*}[-0.1em]{Dice}
        & DSNT & \textbf{0.872} & \textbf{0.884} & \textbf{0.886} & \textbf{0.887} \\
        & & FC   & 0.861 & 0.870 & 0.873 & 0.876\\
        \cmidrule(l){2-7}
        & \multirow{2}{*}[-0.1em]{Haus}
        & DSNT & \textbf{21.6} & 20.07 & \textbf{20.11} & \textbf{19.8} \\
        & & FC   & 21.7 & \textbf{19.91} & 21.03 & 20.3\\
        \midrule
        \multirow{4}{*}[-0.1em]{RIM\_CUP} & \multirow{2}{*}[-0.1em]{Dice}
        & DSNT & \textbf{0.737} & \textbf{0.762} & \textbf{0.768} & \textbf{0.772} \\
        & & FC   & 0.735 & 0.756 & 0.757 & 0.761 \\
        \cmidrule(l){2-7}
        & \multirow{2}{*}[-0.1em]{Haus}
        & DSNT & 19.16 & \textbf{18.62} & \textbf{18.39} & \textbf{18.47} \\
        & & FC   &\textbf{18.46} & 19.04 & 18.59 & 19.09 \\
        \midrule 
        \multirow{4}{*}[-0.1em]{RIM\_DISC} & \multirow{2}{*}[-0.1em]{Dice}
        & DSNT & \textbf{0.949} & \textbf{0.953} & \textbf{0.956} & \textbf{0.957} \\
        & & FC   & 0.949 & 0.951 & 0.951 & 0.953 \\
        \cmidrule(l){2-7}
        & \multirow{2}{*}[-0.1em]{Haus}
        & DSNT & \textbf{10.0} & \textbf{9.45} & \textbf{9.42} & \textbf{9.46} \\
        & & FC   & 10.4 & 10.23 & 10.56 & 10.16 \\
        \bottomrule
    \end{tabular}}
    \caption{Dice and Hausdorff metrics of our model with DSNT or FC head.}
    \label{table:ablation_DSNT}            
\end{table}

\subsubsection{Impact of JS Divergence}
        
We study the effect of the Jensen-Shannon divergence regularization on our model by removing the regularization or by altering $\sigma$ in the covariance $\sigma I_2$ of the 2d Gaussian PDF. As seen from table~\ref{tab:JC_div}, the introduction of the regularization greatly improves model performance, but our model is not sensitive to the choice of $\sigma$.
        
\section{Limitation and Future Works}

There are a couple future research directions that can make the proposed FCSN more robust.

\subsubsection{3D shape learning}

MRI and CT scans are 3D in nature. To apply the current FCSN structure to 3D segmentation tasks, the 3D scan must be interpreted as independent slices. However, the independent assumption across the slices could lead to an inconsistent mask prediction. As a solution to this, one can generalize our framework by modifying our 2D F-DSNT module to a 3D version of it.

\begin{table}[t!]
    \centering
    \resizebox{\columnwidth}{!}{
    \begin{tabular}{lcccc}
        \toprule
        & $\sigma=0.005$ & $\sigma=0.01$ & $\sigma=0.015$ & no JS \\
        \midrule
        Dice & 0.887          & 0.884         & 0.886 & 0.843 \\
        \bottomrule
        \end{tabular}}
        \caption{Dice on ISIC for our model with various regularisation.}
        \label{tab:JC_div}
    \end{table}

\begin{figure}[t!]
    \centering
    \includegraphics[width=7 cm]{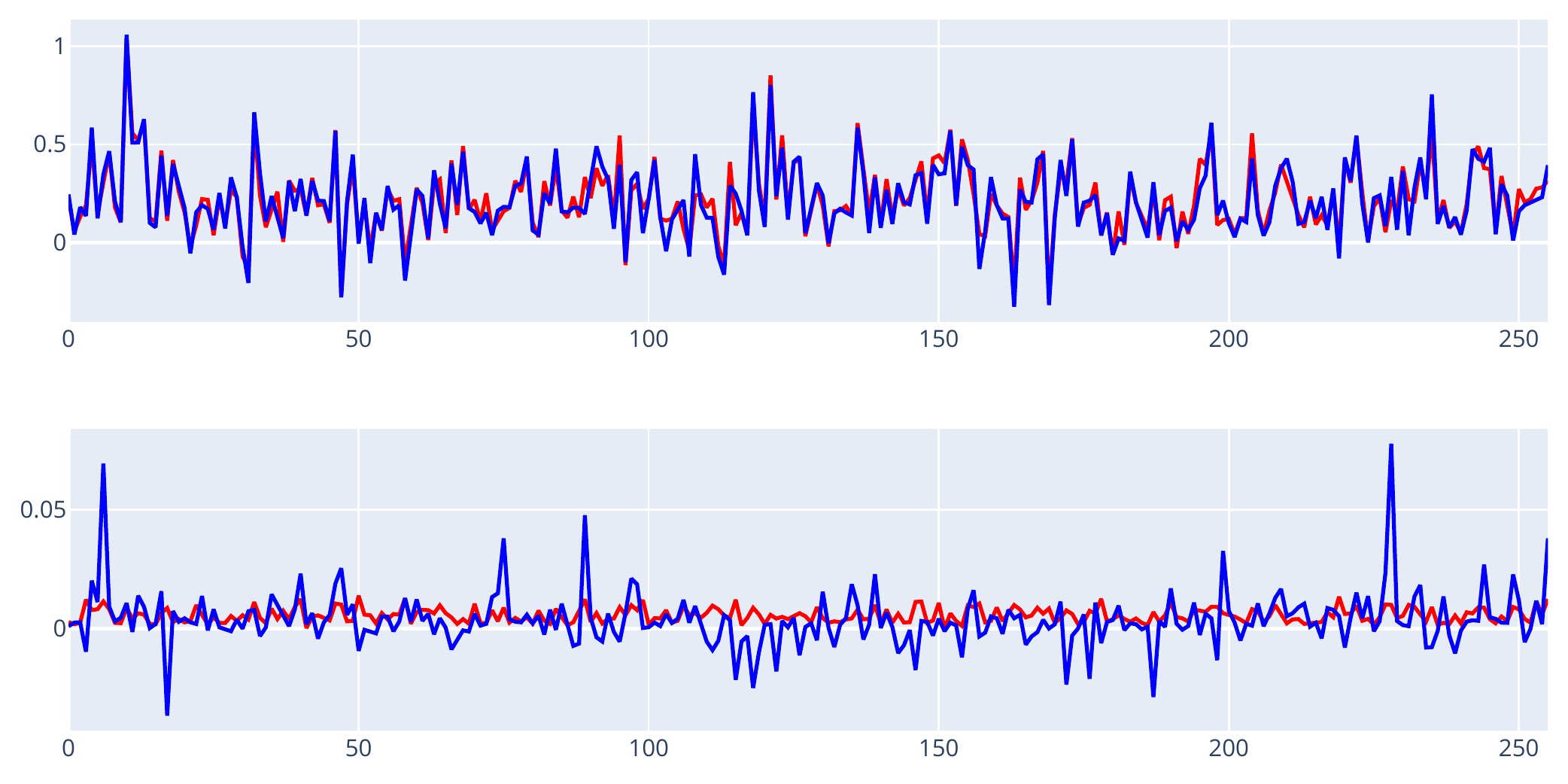}
    \caption{Plots of real parts of Fourier coefficients for a batch of $256$ images. {\color{blue}Blue lines}: Predictions by FCSN. {\color{red} Red lines}: Ground truth. \textbf{Upper}: First Fourier Coefficients. \textbf{Lower}: 10th Fourier Coefficients.}
    \label{fig:High_variance_higher_Fourier_Coef}
\end{figure}

\subsubsection{High Variance of Higher Frequency Coeficients}

Figure~\ref{fig:High_variance_higher_Fourier_Coef} shows that FCSN can give accurate predictions for the first Fourier coefficients, but there are greater mismatch for the higher frequency coefficients. One way to tackle this problem is to introduce multi-heads with various resolutions, where the high resolution head can promote learning of higher Fourier coefficients.

\subsubsection{Multi-object Segmentation Task}

To extend FCSN to multi-instance segmentation cases such as multi-organ segmentation, one could regard FCSN as a segmentation head in MaskRCNN~\cite{he2017mask}.

\subsubsection{Learning other transforms}
FCSN learns to predict Fourier Coefficients for segmentation, and it works well for targets with smooth boundary. However, if the target boundaries contain sharp corners, one may consider modifying FCSN to learn coefficients from more general transforms, like wavelet or tight frame transform. The idea is to use a proper family of base functions that are more efficient in coding boundaries curves. 

\section{Closing Remarks}

In this paper we propose FCSN, a novel and lightweight segmentation model that segments an object by predicting the Fourier coefficient of the object's contour. Our model is designed to incorporate the global context of an image, leading to more accurate segmentation that better preserve the shape and topology of the object. Moreover, the global context awareness makes our model robust to unseen local perturbations during inference.

Our approach is the first step towards a systematic study of performing segmentation by predicting coefficients of mask decomposition. There are many other approaches besides predicting Fourier coefficients. For instance, one can use wavelet or tight frame transforms to obtain more efficient decomposition for boundary curves with sharp corners.

\bibliographystyle{IEEEtran}
\bibliography{ref}

\begin{thebibliography}{10}
\providecommand{\url}[1]{#1}
\csname url@samestyle\endcsname
\providecommand{\newblock}{\relax}
\providecommand{\bibinfo}[2]{#2}
\providecommand{\BIBentrySTDinterwordspacing}{\spaceskip=0pt\relax}
\providecommand{\BIBentryALTinterwordstretchfactor}{4}
\providecommand{\BIBentryALTinterwordspacing}{\spaceskip=\fontdimen2\font plus
\BIBentryALTinterwordstretchfactor\fontdimen3\font minus
  \fontdimen4\font\relax}
\providecommand{\BIBforeignlanguage}[2]{{%
\expandafter\ifx\csname l@#1\endcsname\relax
\typeout{** WARNING: IEEEtran.bst: No hyphenation pattern has been}%
\typeout{** loaded for the language `#1'. Using the pattern for}%
\typeout{** the default language instead.}%
\else
\language=\csname l@#1\endcsname
\fi
#2}}
\providecommand{\BIBdecl}{\relax}
\BIBdecl

\bibitem{ronneberger2015u}
O.~Ronneberger, P.~Fischer, and T.~Brox, ``U-net: Convolutional networks for
  biomedical image segmentation,'' in \emph{International Conference on Medical
  image computing and computer-assisted intervention}.\hskip 1em plus 0.5em
  minus 0.4em\relax Springer, 2015, pp. 234--241.

\bibitem{chen2018encoder}
L.-C. Chen, Y.~Zhu, G.~Papandreou, F.~Schroff, and H.~Adam, ``Encoder-decoder
  with atrous separable convolution for semantic image segmentation,'' in
  \emph{Proceedings of the European conference on computer vision (ECCV)},
  2018, pp. 801--818.

\bibitem{dosovitskiy2020image}
A.~Dosovitskiy, L.~Beyer, A.~Kolesnikov, D.~Weissenborn, X.~Zhai,
  T.~Unterthiner, M.~Dehghani, M.~Minderer, G.~Heigold, S.~Gelly \emph{et~al.},
  ``An image is worth 16x16 words: Transformers for image recognition at
  scale,'' \emph{arXiv preprint arXiv:2010.11929}, 2020.

\bibitem{liu2015parsenet}
W.~Liu, A.~Rabinovich, and A.~C. Berg, ``Parsenet: Looking wider to see
  better,'' \emph{arXiv preprint arXiv:1506.04579}, 2015.

\bibitem{farabet2012learning}
C.~Farabet, C.~Couprie, L.~Najman, and Y.~LeCun, ``Learning hierarchical
  features for scene labeling,'' \emph{IEEE transactions on pattern analysis
  and machine intelligence}, vol.~35, no.~8, pp. 1915--1929, 2012.

\bibitem{geirhos2020shortcut}
R.~Geirhos, J.-H. Jacobsen, C.~Michaelis, R.~Zemel, W.~Brendel, M.~Bethge, and
  F.~A. Wichmann, ``Shortcut learning in deep neural networks,'' \emph{Nature
  Machine Intelligence}, vol.~2, no.~11, pp. 665--673, 2020.

\bibitem{nibali2018numerical}
A.~Nibali, Z.~He, S.~Morgan, and L.~Prendergast, ``Numerical coordinate
  regression with convolutional neural networks,'' \emph{arXiv preprint
  arXiv:1801.07372}, 2018.

\bibitem{long2015fully}
J.~Long, E.~Shelhamer, and T.~Darrell, ``Fully convolutional networks for
  semantic segmentation,'' in \emph{Proceedings of the IEEE conference on
  computer vision and pattern recognition}, 2015, pp. 3431--3440.

\bibitem{wang2020non}
Z.~Wang, N.~Zou, D.~Shen, and S.~Ji, ``Non-local u-nets for biomedical image
  segmentation,'' in \emph{Proceedings of the AAAI Conference on Artificial
  Intelligence}, vol.~34, no.~04, 2020, pp. 6315--6322.

\bibitem{chen2021transunet}
J.~Chen, Y.~Lu, Q.~Yu, X.~Luo, E.~Adeli, Y.~Wang, L.~Lu, A.~L. Yuille, and
  Y.~Zhou, ``Transunet: Transformers make strong encoders for medical image
  segmentation,'' \emph{arXiv preprint arXiv:2102.04306}, 2021.

\bibitem{vaswani2017attention}
A.~Vaswani, N.~Shazeer, N.~Parmar, J.~Uszkoreit, L.~Jones, A.~N. Gomez,
  {\L}.~Kaiser, and I.~Polosukhin, ``Attention is all you need,'' in
  \emph{Advances in neural information processing systems}, 2017, pp.
  5998--6008.

\bibitem{luo2016understanding}
W.~Luo, Y.~Li, R.~Urtasun, and R.~Zemel, ``Understanding the effective
  receptive field in deep convolutional neural networks,'' \emph{Advances in
  neural information processing systems}, vol.~29, 2016.

\bibitem{berman2018lovasz}
M.~Berman, A.~R. Triki, and M.~B. Blaschko, ``The lov{\'a}sz-softmax loss: A
  tractable surrogate for the optimization of the intersection-over-union
  measure in neural networks,'' in \emph{Proceedings of the IEEE conference on
  computer vision and pattern recognition}, 2018, pp. 4413--4421.

\bibitem{oktay2017anatomically}
O.~Oktay, E.~Ferrante, K.~Kamnitsas, M.~Heinrich, W.~Bai, J.~Caballero, S.~A.
  Cook, A.~De~Marvao, T.~Dawes, D.~P. O‘Regan \emph{et~al.}, ``Anatomically
  constrained neural networks (acnns): application to cardiac image enhancement
  and segmentation,'' \emph{IEEE transactions on medical imaging}, vol.~37,
  no.~2, pp. 384--395, 2017.

\bibitem{jia2021regularized}
F.~Jia, J.~Liu, and X.-C. Tai, ``A regularized convolutional neural network for
  semantic image segmentation,'' \emph{Analysis and Applications}, vol.~19,
  no.~01, pp. 147--165, 2021.

\bibitem{jia2020nonlocal}
F.~Jia, X.-C. Tai, and J.~Liu, ``Nonlocal regularized cnn for image
  segmentation,'' \emph{Inverse Problems \& Imaging}, vol.~14, no.~5, p. 891,
  2020.

\bibitem{liu2020abcnet}
Y.~Liu, H.~Chen, C.~Shen, T.~He, L.~Jin, and L.~Wang, ``Abcnet: Real-time scene
  text spotting with adaptive bezier-curve network,'' in \emph{Proceedings of
  the IEEE/CVF Conference on Computer Vision and Pattern Recognition}, 2020,
  pp. 9809--9818.

\bibitem{chen2021bezierseg}
H.~Chen, Y.~Deng, B.~Li, Z.~Li, H.~Chen, B.~Jing, and C.~Li, ``Bezierseg:
  Parametric shape representation for fast object segmentation in medical
  images,'' \emph{arXiv preprint arXiv:2108.00760}, 2021.

\bibitem{xie2020polarmask}
E.~Xie, P.~Sun, X.~Song, W.~Wang, X.~Liu, D.~Liang, C.~Shen, and P.~Luo,
  ``Polarmask: Single shot instance segmentation with polar representation,''
  in \emph{Proceedings of the IEEE/CVF conference on computer vision and
  pattern recognition}, 2020, pp. 12\,193--12\,202.

\bibitem{riaz2021fouriernet}
H.~U.~M. Riaz, N.~Benbarka, and A.~Zell, ``Fouriernet: Compact mask
  representation for instance segmentation using differentiable shape
  decoders,'' in \emph{2020 25th International Conference on Pattern
  Recognition (ICPR)}.\hskip 1em plus 0.5em minus 0.4em\relax IEEE, 2021, pp.
  7833--7840.

\bibitem{codella2019skin}
N.~Codella, V.~Rotemberg, P.~Tschandl, M.~E. Celebi, S.~Dusza, D.~Gutman,
  B.~Helba, A.~Kalloo, K.~Liopyris, M.~Marchetti \emph{et~al.}, ``Skin lesion
  analysis toward melanoma detection 2018: A challenge hosted by the
  international skin imaging collaboration (isic),'' \emph{arXiv preprint
  arXiv:1902.03368}, 2019.

\bibitem{RIMONEDLImageAnalStereol2346}
\BIBentryALTinterwordspacing
F.~J.~F. Batista, T.~Diaz-Aleman, J.~Sigut, S.~Alayon, R.~Arnay, and
  D.~Angel-Pereira, ``Rim-one dl: A unified retinal image database for
  assessing glaucoma using deep learning,'' \emph{Image Analysis \&
  Stereology}, vol.~39, no.~3, pp. 161--167, 2020. [Online]. Available:
  \url{https://www.ias-iss.org/ojs/IAS/article/view/2346}
\BIBentrySTDinterwordspacing

\bibitem{kingma2014adam}
D.~P. Kingma and J.~Ba, ``Adam: A method for stochastic optimization,''
  \emph{arXiv preprint arXiv:1412.6980}, 2014.

\bibitem{zhou1807nested}
Z.~Zhou, M.~Siddiquee, N.~Tajbakhsh, and J.~Liang, ``A nested u-net
  architecture for medical image segmentation,'' \emph{arXiv preprint
  arXiv:1807.10165}, 2018.

\bibitem{hatamizadeh2022unetr}
A.~Hatamizadeh, Y.~Tang, V.~Nath, D.~Yang, A.~Myronenko, B.~Landman, H.~R.
  Roth, and D.~Xu, ``Unetr: Transformers for 3d medical image segmentation,''
  in \emph{Proceedings of the IEEE/CVF Winter Conference on Applications of
  Computer Vision}, 2022, pp. 574--584.

\bibitem{yu2017dilated}
F.~Yu, V.~Koltun, and T.~Funkhouser, ``Dilated residual networks,'' in
  \emph{Proceedings of the IEEE conference on computer vision and pattern
  recognition}, 2017, pp. 472--480.

\bibitem{he2017mask}
K.~He, G.~Gkioxari, P.~Doll{\'a}r, and R.~Girshick, ``Mask r-cnn,'' in
  \emph{Proceedings of the IEEE international conference on computer vision},
  2017, pp. 2961--2969.

\end{thebibliography}

\end{document}